\documentclass[conference]{IEEEtran}
\IEEEoverridecommandlockouts
\usepackage{cite}
\usepackage{amsmath,amssymb,amsfonts}
\usepackage{algorithm}
\usepackage{algorithmicx}
\usepackage{algpseudocode}
\usepackage{graphicx}
\usepackage{textcomp}
\usepackage{xcolor}

\newtheorem{Theorem}{Theorem}

\usepackage{mathtools,enumitem,color,subfigure,bbm}
\usepackage{graphicx,tabularx,array}
\usepackage{hyperref}
\DeclareMathOperator*{\argmax}{arg\,max}

\begin{document}

\title{
Deep Reinforcement Learning for MIMO Communication with Low-Resolution ADCs
}

\author{\IEEEauthorblockN{Marian Temprana Alonso}
\IEEEauthorblockA{\textit{Knight Foundation School of}\\ \textit{Computing and Information Sciences} \\
\textit{Florida International University}\\
Miami, United States \\
mtemp009@fiu.edu}
\and
\IEEEauthorblockN{Dongsheng Luo}
\IEEEauthorblockA{\textit{Knight Foundation School of}\\ \textit{Computing and Information Sciences} \\
\textit{Florida International University}\\
Miami, United States \\
dluo@fiu.edu}
\and
\IEEEauthorblockN{Farhad Shirani}
\IEEEauthorblockA{\textit{Knight Foundation School of}\\ \textit{Computing and Information Sciences} \\
\textit{Florida International University}\\
Miami, United States \\
fshirani@fiu.edu}
}

\maketitle

\begin{abstract}
Multiple-input multiple-output (MIMO) wireless systems conventionally use high-resolution analog-to-digital converters (ADCs) at the receiver side to faithfully digitize received signals prior to digital signal processing. However, the power consumption of ADCs increases significantly as the bandwidth is increased, particularly in millimeter wave communications systems. A combination of two mitigating approaches has been considered in the literature: i) to use hybrid beamforming to reduce the number of ADCs, and ii) to use low-resolution ADCs to reduce per ADC power consumption.
Lowering the number and resolution of the ADCs naturally reduces the communication rate of the system, leading to a tradeoff between ADC power consumption and communication rate. Prior works have shown that optimizing over the hybrid beamforming matrix and ADC thresholds may reduce the aforementioned rate-loss significantly. A key challenge is the complexity of optimization over all choices of beamforming matrices and threshold vectors. This work proposes a reinforcement learning (RL) architecture to perform the optimization. The proposed approach integrates deep neural network-based mutual information estimators for reward calculation with policy gradient methods for reinforcement learning. The approach is robust to dynamic channel statistics and noisy CSI estimates. It is shown theoretically that greedy RL methods converge to the globally optimal policy. Extensive empirical evaluations are provided demonstrating that the performance of the RL-based approach closely matches exhaustive search optimization across the solution space.
\footnote{To facilitate reproducibility, the code associated with this work is available at 
\url{https://github.com/mtalonso-research/RL_MIMO.git}}
\end{abstract}

\section{Introduction}
Millimeter wave (mm-wave) communication systems have emerged as a key technology for enabling high data rate wireless communications due to the abundant spectrum available at frequencies above 6 GHz. In mm-wave cellular applications, bandwidths above 500 MHz are considered (e.g., 3GPP 5G NR \cite{3GPP_TS38_104}), compared to 1.4-20 MHz in LTE protocols. However, the increased bandwidth presents significant challenges related to power consumption, particularly in the analog-to-digital converters (ADCs) and digital-to-analog converters (DACs) whose power consumption scales with bandwidth \cite{razavi2005design}.
The power consumption issue is further exacerbated in multiple-input multiple-output (MIMO) systems, which employ large antenna arrays to mitigate the path loss experienced at high carrier frequencies. In conventional MIMO systems with digital beamforming, each receive antenna requires a dedicated ADC, resulting in substantial power consumption that is incompatible with the limited energy budget of mobile devices and small-cell access points. For instance, current commercial high-speed ($\geq$ 20 GSample/s), high-resolution (8-12 bits) ADCs consume approximately 500 mW per converter \cite{zhang2018low}.

Two mitigating approaches have been proposed to address the ADC power consumption. The first approach is to use analog or hybrid beamforming architectures to reduce the number of required ADCs \cite{molisch2017hybrid,heath2016overview,alkhateeb2014mimo,zirtiloglu2022}. The second approach utilizes low-resolution ADCs (1-3 bits) to decrease the power consumption per converter \cite{miao2016joint,shirani2022mimo}. However, both approaches introduce performance penalties due to coarse quantization and/or reduced spatial multiplexing capabilities. Prior works have demonstrated that careful selection of the hybrid beamforming matrix and ADC threshold levels can significantly mitigate these performance losses \cite{mo2015capacity,khalili2021mimo}. However, jointly optimizing these parameters is computationally complex due to the high-dimensional and non-convex nature of the problem. In this work, we propose a deep reinforcement learning (RL) based method to solve this optimization problem.

Machine learning (ML) techniques have gained significant traction in addressing complex optimization problems in wireless communications. For quantization-related problems, \cite{shlezinger2021deep,mashhadi2020distributed,liang2022changeable} have demonstrated the ability to learn quantizers directly from data. Similarly, ML-based approaches for beamforming optimization have shown promising results \cite{eldar2022machine,heng2021machine}. In the context of reinforcement learning (RL) for communications, several recent works have explored policy-based optimization for resource allocation \cite{naderializadeh2021resource,ma2024efficient}. A critical component in RL is the reward estimation, which often requires mutual information computation in communication systems. Several neural estimators for mutual information have been proposed, including CORTICAL \cite{letizia2022}, MINE \cite{belghazi2018}, SMILE \cite{song2019understanding}, and MMIE \cite{li2024}, among others, which can be leveraged for accurate reward computation during RL training. These neural estimators enable end-to-end optimization of communication systems without relying on simplified analytical expressions that may not capture the true performance in practical scenarios with complex channel models and hardware characteristics.

In this work, we introduce an RL architecture to jointly optimize beamforming matrices and ADC threshold levels in mm-wave MIMO systems with low-resolution ADCs. To our knowledge, this is the first instance of applying RL techniques for receiver-side optimization in MIMO systems with low-resolution ADC quantization.  
Our contributions include:
\begin{itemize}
\item We formulate the joint optimization of beamforming matrices and ADC threshold vectors as an RL problem with the objective of maximizing the achievable communication rate under power constraints.
\item We develop an approach that integrates neural network-based mutual information estimators (specifically utilizing CORTICAL \cite{letizia2022}) for accurate reward calculation with policy gradient methods.
\item We provide theoretical analysis proving the convergence of the proposed RL method to globally optimal solutions under mild assumptions.
\item We demonstrate through extensive simulations that our approach achieves performance comparable to exhaustive search optimization and with significant reduction in computational complexity.
\end{itemize}

% The main contributions of this work are summarized below:
% \begin{itemize}
%     \item To define and implement a framework for selecting optimal quantization thresholds and reconstruction points for a quantization setup.
%     \item To implement this framework using the Blahut-Arimoto algorithm and CORTICAL as mutual information estimators.
%     \item To establish theoretical guarantees for the convergence of the proposed framework to the optimal quantization configuration.
%     \item To empirically evaluate the performance of the framework in both hybrid and analog beamforming settings.
% \end{itemize}

\noindent {\em Notation:}
%Sets are denoted by calligraphic letters such as $\mathcal{X}$.
%, families of sets by sans-serif letters such as $\mathsf{X}$. 
The set $\{1,2,\cdots, n\}$ is represented by $[n]$. 
%$\mathcal{X}^c$ denotes the complement of $\mathcal{X}$.
The vector $(x_1,x_2,\hdots, x_n)$ is written as $x^n$, and 
%$(x_k,x_{k+1},\cdots,x_n)$ is denoted by $x_{k}^n$. 
the $i$th element is written as $x_i$.  An $n\times m$ matrix is written as $h^{n\times m}=[h_{i,j}]_{i\in [n], j\in [m]}$.
%, its $i$th row is denoted by $h_{i,:}=[h_{i,j}]_{j\in [m]}$ and its $j$th column is $h_{:,j}=[h_{i,j}]_{i\in [n]}$. 
The $n\times n$ identity matrix is denoted by $\mathbf{I}_n$.
We use bold-face letters such as $\mathbf{x}$ and $\mathbf{h}$ instead of $x^n$ and $h^{n\times m}$, respectively, when the dimension is clear from the context.  
%$\mathbf{x}^H$ denotes the hermitian of $\mathbf{x}$. 
We write $||\cdot||_2$ to denote the $L_2$-norm.  
Upper-case letters represent random variables, and lower-case letters represent their realizations. For a Gaussian random vector $\mathbf{X}$ 
with mean vector $\pmb{\mu}$ and covariance matrix $\mathbf{\Sigma}$, we write $\mathbf{X}\sim\mathcal{N}( \pmb{\mu},\mathbf{\Sigma})$. 

\section{Preliminaries}
\label{sec:Prelim}
\subsection{Problem Formulation}
\label{sec:PF}
\begin{figure}[t] 
    \centering    \includegraphics[width=0.49\textwidth]{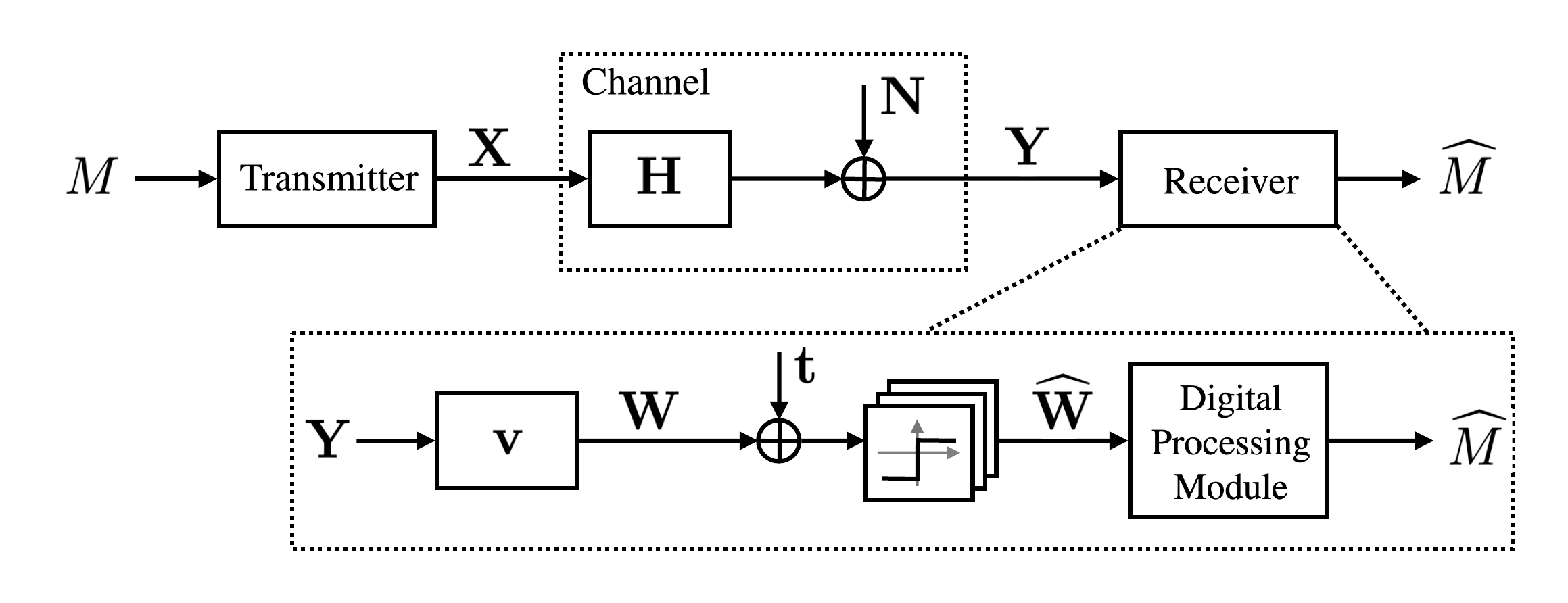}
    \vspace{-.2in}
    \caption{Overview of the MIMO communication system. }
    \label{fig:system_model}
    \vspace{-.2in}
\end{figure}

We consider a MIMO communication system consisting of \( n_t \) transmit antennas and $n_r$ receive antennas (Figure \ref{fig:system_model}). The message ${M}$ is mapped to\footnote{We note that this mapping is blockwise. However, in the problem formulation, we focus on a single channel-use to simplify notation.} the channel input vector $\mathbf{X} \in \mathbb{R}^{n_t}$, which is subject to average power constraint $\mathbb{E}(||\mathbf{X}||_2^2) \leq P_T$. 
The channel output vector $\mathbf{Y} \in \mathbb{R}^{n_r}$ is given as
\begin{align*}
    \mathbf{Y} = \mathbf{H} \mathbf{X} + \mathbf{N},
\end{align*}
where $\mathbf{H} \in \mathbb{R}^{n_r \times n_t}$ is the channel gain matrix and $\mathbf{N} \sim \mathcal{N}(0, \mathbf{I}_{n_r})$ is a vector of independent Gaussian variables with zeo mean and unit variance.
The signal $\mathbf{Y}$ is first processed by an analog linear combiner $\mathbf{v} \in \mathbb{R}^{n_q \times n_r}$, where $n_q$ denotes the number of ADCs, i.e., the combiner outputs $\mathbf{W} = \mathbf{v}\mathbf{Y}$. Each component $W_i$ is passed through a dedicated ADC, yielding the quantized signal $\widehat{{W}_i} \in \{0,1,\cdots,\ell-1\}$, where $\ell$ is the number of ADC levels. The operation of the ADC is detailed in the sequel.
The resulting discretized vector $\widehat{\mathbf{W}}$ is then passed to a (digital) blockwise processing module that decodes the reconstructed message $\widehat{M}$. 

% The ADCs are represented as a vector of quantizers $\mathbf{q}(\cdot) = (q_1(\cdot),q_2(\cdot)\cdots q_{n_q}(\cdot))$, applied element-wise to each $W_i$. Each quantizer is defined by a threshold vector $\mathbf{t} = (t_1, t_2, \dots, t_{n_q})$ and corresponding reconstruction points $\mathbf{c}\in\mathbb{R}^{\ell\times n_r}$, where $\ell$ corresponds to the number of reconstruction points for a given configuration, defined as $\ell = \sum^{n_r}_{k=0}\binom{n_q}{k}$. 
% %\textcolor{red}{I am not sure if it makes sense to keep going with the $n_r$ or if it would make more sense to use $n_s$ as the number of RF chains from $n_r$.} 
% Then, given thresholds $t_1 < t_2 < \dots < t_{n_q}$, each $w$ is quantized as follows:
% \begin{align*}
%     \widehat{w} = \mathbf{q}(w) = 
%     \begin{cases}
%     c_0, & \text{if } w < t_1 \\
%     c_i, & \text{if } t_i \leq w < t_{i+1}, \quad i \in \{1, \dots, n_q\} \\
%     c_{\ell}, & \text{if } t_{n_q} \leq w
%     \end{cases}
% \end{align*}
%\textcolor{red}{verify this for multidimensional case}

We assume that each of the ADCs have $\ell$ output levels.  An ADC is represented by a mapping from  a continuous-valued $w$ to a discrete-valued $\widehat{w}$. Given the threshold vector $t^{\ell-1}=(t_1,t_2,\cdots,t_{\ell-1})$, The output 
of the ADC
is given by:
 \begin{align}
     \widehat{w}= 
     %q(y)=
     \begin{cases}
         0 \qquad &\text{if } \quad  w<t_1\\
         i &\text{if } \quad \exists i\in [\ell-2], t_i\leq w < t_{i+1}\\
         \ell-1 &\text{if } \quad t_{\ell-1}\leq w 
     \end{cases},
     \label{eq:ADC}
 \end{align}
 where $t_1<t_2<\cdots<t_{\ell-1}$.
 For a receiver equipped with $n_q$ ADCs each with $\ell$ output levels, the threshold matrix is defined as $\mathbf{t}\in \mathbb{R}^{n_q\times {(\ell-1)}}$, where $t_{i,:}, i\in [n_q]$ is the threshold vector corresponding to the $i$th ADC. %The quantization function corresponding to the $i$th ADC is denoted by $q_i(\cdot)$. The ADC module is represented by $\mathbf{a}(\cdot)= (q_1(\cdot),q_2(\cdot),\cdots, q_{n_q}(\cdot))$.

Note that since the channel is discrete-output, the input alphabet can be restricted to a discrete set \cite{witsenhausen1980some,singh2009limits}. Consequently, for a given CSI matrix $\mathbf{H}$, linear combiner $\mathbf{v}$, and threshold vector $\mathbf{t}$, the channel capacity is characterized as:
\begin{align*}
  C=\max_{\mathcal{X}} \max_{P_{\mathbf{X}}} I(\mathbf{X};\mathbf{\widehat{W}}),
\end{align*}
where $\mathcal{X}=\{c_1,c_2,\cdots,c_{\xi}\}$ is the channel input alphabet, $c_i\in \mathbb{R}^{n_r}$, $\xi$ is the maximum number of quantization regions, and its value is characterized in terms of the parameters\footnote{It follows from  \cite[Corollary 1]{winder1966partitions} that $\xi\leq 2\sum_{i=0}^{n_r-1}{(\ell-1) n_q-1 \choose i}.$
} $(n_r,n_q,\ell)$, and $P_{\mathbf{X}}$ is a probability distribution over $\mathcal{X}$, satisfying the input power constraint $\mathbb{E}(||\mathbf{X}||_2^2) \leq P_T$.

Given CSI matrix $\mathbf{H}$ (or a CSI estimate $\widehat{\mathbf{H}}$), our objective is to find $(\mathbf{v}^*,\mathbf{t}^*,\mathcal{X}^*)$ such that the resulting channel capacity $C$ is maximized, i.e., to find
\begin{align*}
(\mathbf{v}^*,\mathbf{t}^*,\mathcal{X}^*)=\arg\max_{(\mathbf{v},\mathbf{t}, \mathcal{X})}\max_{P_{\mathbf{X}}} I(\mathbf{X};\mathbf{\widehat{W}}).
\end{align*}
We introduce an RL framework towards this objective.
\subsection{The Policy Gradients Method}
An RL problem is typically formulated as a Markov Decision Process (MDP), defined by a tuple \((\mathcal{S}, \mathcal{A}, P, R, \gamma)\), where \(\mathcal{S}\) and \(\mathcal{A}\) denote the state and action spaces, respectively, \(P\) represents the state-transition probability, \(R\) is the reward function, and \(\gamma \in [0,1]\) is a discount factor \cite{sutton2018reinforcement}. The policy gradients method is an RL technique used to optimize parametric policies (e.g., the policy which finds $(\mathbf{v}^*,\mathbf{t}^*,\mathcal{X}^*)$) by directly approximating the gradient of the expected reward (e.g., resulting channel capacity) with respect to policy parameters. Formally, consider a policy \(\pi_{\theta}(a|s)\) parameterized by \(\theta\),
mapping any given state realization \(s\) to a probability distribution over actions \(a\). The objective is to maximize the expected cumulative reward defined by:
\[
J(\theta) = \mathbb{E}_{\pi_{\theta}}\left[\sum_{i=0}^{T} \gamma^{i} R(S_i, A_i)\right],
\]
where \(R(S_i,A_i)\) is the reward obtained at time step \(i\), and the pair of random variables $(S_i,A_i)$ are the state-action pair at time $i$ whose underlying probability distribution depends on the policy $\pi_{\theta}$ and the state transition probability $P$. The gradient with respect to policy parameters \(\theta\) is given by:
\[
\nabla_{\theta} J(\theta) = \mathbb{E}_{\pi_{\theta}}\left[\sum_{i=0}^{T}\gamma^{i}\nabla_{\theta}\log\pi_{\theta}(A_i|S_i)Q^{\pi_{\theta}}(S_i,A_i)\right].
\]
where \(Q^{\pi_{\theta}}(\cdot,\cdot)\) is the state-action value function:
\[
Q^{\pi_{\theta}}(s,a)=\mathbb{E}_{\pi_{\theta}}\left[\sum_{i'=i}^{T}\gamma^{i'-i} R(S_{i'},A_{i'})\middle| S_i=s,A_i=a\right],
\]
representing the expected cumulative reward from taking action \(a\) in state \(s\) and thereafter following policy \(\pi_{\theta}\). The interested reader may refer to \cite{sutton2018reinforcement} for a comprehensive description of the policy gradients method.

\section{Reinforcement Learning for Receiver-side Optimization}

\begin{figure}[t] 
    \centering
    \includegraphics[width=\columnwidth]{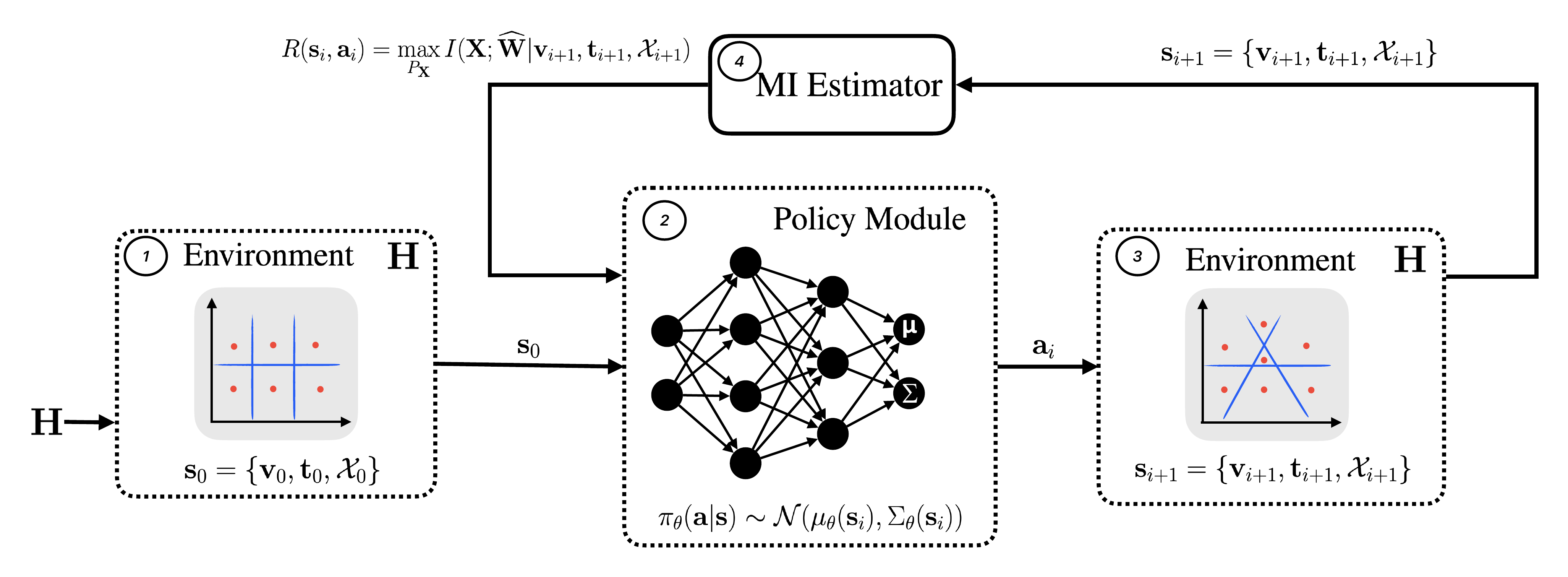}
    \caption{Overview of the proposed reinforcement learning model, where the environment is defined by $\mathbf{H}$ and the initial state consisting of analog processing matrix $\mathbf{v}$, thresholds $\mathbf{t}$, and input alphabet $\mathcal{X}$. The policy then takes the current state as input and outputs a distribution with mean $\mu$ and covariance $\Sigma$ from which an action is chosen to determine the next state. Finally, the reward is computed by a mutual information estimator.}
    \label{fig:rl_overview}
\end{figure}

In this section, we propose an RL framework to optimize the receiver configuration in the MIMO system described in Section \ref{sec:PF}. Specifically, our objective is to find the analog linear combiner matrix $\mathbf{v}$, ADC threshold matrix $\mathbf{t}$, and input alphabet $\mathcal{X}$ that maximize the channel capacity characterized by the mutual information $\max_{P_{\mathbf{X}}} I(\mathbf{X}; \widehat{\mathbf{W}})$. Due to the complex, non-convex nature of this optimization problem, particularly the joint optimization of continuous parameters $(\mathbf{v}, \mathbf{t},\mathcal{X})$ and the input distribution $P_{\mathbf{X}}$, we employ the REINFORCE policy gradient algorithm \cite{sutton1999reinforcement} with an additional Kullback-Leibler (KL) divergence penalty term inspired by proximal policy optimization (PPO) \cite{schulman2017proximal} as described in the following.

\subsection{Markov Decision Process Formulation}
\label{sec:MDP}
We model the optimization process as an MDP. The components of our MDP are defined as follows:
\\\textbf{State Space ($\mathcal{S}$):} The state at step $i$, denoted by $s_i \in \mathcal{S}$, represents the current configuration of the receiver parameters being optimized:
    \[
    s_i = (\mathbf{v}_i, \mathbf{t}_i,\mathcal{X}_i) \in \mathbb{R}^{n_q \times n_r} \times \mathbb{R}^{n_q \times (\ell-1)}\times \mathbb{R}^{\xi}.
    \]
    Note that the channel matrix $\mathbf{H}$ is considered part of the environment dynamics rather than the state.
    \\\textbf{Action Space ($\mathcal{A}$):} The action $a_i \in \mathcal{A}$ represents the adjustments made to the receiver parameters at step $i$. The action $a_i$ consists of additive updates to the current parameters:
    \[
    a_i = (\Delta \mathbf{v}_i, \Delta \mathbf{t}_i,\Delta \mathcal{X}_i),
    \]
    where $\Delta \mathbf{v}_i \in \mathbb{R}^{n_q \times n_r}$, $\Delta \mathbf{t}_i \in \mathbb{R}^{n_q \times (\ell-1)}$, and $\Delta \mathcal{X}_i\in \mathbb{R}^\xi$ are the adjustments sampled from the agent's policy.
\\\textbf{Policy ($\pi_{\theta}$):} The agent's actions are governed by a stochastic policy $\pi_{\theta}$, mapping the current state $S_i$ to a  distribution over action space $\mathcal{A}$. We employ a Gaussian policy:
    \[
    A_i \sim \mathcal{N}(\pmb{\mu}_{\theta}(S_i), \mathbf{\Sigma}_{\theta}(S_i)).
    \]
    The policy network outputs the mean $\pmb{\mu}_{\theta}(S_i) = (\pmb{\mu}_{\mathbf{v}}(S_i), \pmb{\mu}_{\mathbf{t}}(S_i), \pmb{\mu}_{\mathcal{X}}(S_i))$ and the standard variance $\mathbf{\Sigma}_{\theta}(S_i) = (\mathbf{\Sigma}_{\mathbf{v}}(S_i), \mathbf{\Sigma}_{\mathbf{t}}(S_i),\mathbf{\Sigma}_{\mathcal{X}}(S_i))$.
\\\textbf{State Transition ($P$):} The state transition is deterministic given the action. After applying $A_i$, the next state is:
    \[
    S_{i+1}\! =\! (\mathbf{v}_{i+1}, \mathbf{t}_{i+1},\mathcal{X}_{i+1}) \!=\! (\mathbf{v}_i + \Delta \mathbf{v}_i, \mathbf{t}_i + \Delta \mathbf{t}_i,\mathcal{X}_i+\Delta \mathcal{X}_i).
    \]
    In our implementation, we include a projection step  to ensure threshold order $t_{i,1} \leq \dots \leq t_{i, \ell-1}$.
\\\textbf{Reward Function ($R$):} The reward $R(S_i, A_i)$ is:
    \begin{align*}
      &  R(S_i, A_i)=
      \\& \max_{P_X}I(\mathbf{X}; \widehat{\mathbf{W}} | \mathbf{v}_{i+1}, \mathbf{t}_{i+1},\mathcal{X}_{i+1}) - \lambda_1 \mathcal{L}_{\text{power}},
    \end{align*}
    where $\mathcal{L}_{\text{power}}= |\mathbb{E}_{X}( \| X\|_2^2)-P_T|^+$, and $\lambda_1 \ge 0$ is a hyperparameter.
\subsection{Policy Optimization}
We optimize the policy parameters $\theta$ using the REINFORCE algorithm \cite{sutton1999reinforcement}, augmented with a Kullback-Leibler (KL) divergence penalty term inspired by PPO \cite{schulman2017proximal} to promote training stability. The objective is to maximize the expected sum of discounted rewards, while penalizing large changes in the policy between updates. We define the objective function to be maximized as:
\begin{align*}
J(\theta) =& \mathbb{E}_{\tau \sim \pi_{\theta_{\text{old}}}} \left[ \sum_{i=0}^{T-1} \gamma^i \log \pi_{\theta}(A_i | S_i) R(S_i, A_i) \right] 
\\&- \beta \, \mathbb{E}_{S \sim \tau} \left[ \text{KL}(\pi_{\theta_{\text{old}}}(\cdot | S) \,\|\, \pi_\theta(\cdot | S)) \right],
\end{align*}
where the expectation $\mathbb{E}_{\tau \sim \pi_{\theta_{\text{old}}}}$ is taken over trajectories $\tau = (S_0, A_0, R_0, S_1, \dots)$ sampled using the policy $\pi_{\theta_{\text{old}}}$ from the previous iteration.  Note that this formulation uses the full return as opposed to standard PPO which typically uses an advantage function. The inclusion of the KL penalty term, borrowed from PPO, reduces the policy variations due to noisy gradient estimation. 
The parameters $\theta$ are updated iteratively using gradient ascent on $J(\theta)$.

\subsection{Training Procedure} 
The training procedure is outlined in Algorithm~\ref{alg:training_kl_revised} and shown in Figure \ref{fig:rl_overview}. The algorithm iteratively refines the policy parameters $\theta$ over $N$ main iterations. Within each iteration, the current policy $\pi_{\theta_{\text{old}}}$ is used to collect a batch $\mathcal{D}$ of $M$ trajectories (Lines 1-16). 
For each trajectory, a specific Signal-to-Noise Ratio (SNR) value $\text{SNR}_m$ is first sampled uniformly from the range $[\sigma_{\min}, \sigma_{\max}]$, which in turn determines the transmit power $P_T$. During each trajectory, actions $A_i$ are sampled based on the state $S_i$, leading to a next state $S_{i+1}$, and a reward $R_i$ is calculated by estimating the maximum achievable mutual information $I^*$ (subject to power constraint $P_T$) for the resulting state $S_{i+1}$ under the sampled $\text{SNR}_m$. The algorithm then performs $K$ update epochs. In each epoch, it calculates a gradient estimate $\hat{g}$ by combining a policy gradient term $\hat{g}_{\text{policy}}$ (promoting actions leading to higher rewards) and a KL divergence penalty gradient $\hat{g}_{\text{KL}}$ (regularizing the update size), and then updates the policy parameters $\theta$ using this gradient and the learning rate $\alpha$ (Lines 17-24).

\begin{algorithm}[h] 
\caption{Policy Training}
\label{alg:training_kl_revised}
\begin{algorithmic}[1]
\Require $S_0 = (\mathbf{v}_0, \mathbf{t}_0, \mathcal{X}_0)$, learning rate $\alpha$, KL weight $\beta$, power constraint weight $\lambda_1$, discount factor $\gamma$, number of iterations $N$, steps per trajectory $T$, batch size $M$, update epochs $K$, maximum and minumum SNR $\sigma_{max}, \sigma_{min}$.
\vspace{-.1in}
\Statex
\State Initialize policy weights $\theta$.
\For{j = 1 to $N$}
    \State $\theta_{\text{old}} \gets \theta$,  $\mathcal{D} \gets \emptyset$
    \For{$m = 1$ to $M$} \Comment{Collect batch of trajectories}
     \State $\text{SNR}_m \sim \text{Uniform}(\sigma_{\min}, \sigma_{\max})$
        \State $\tau_m \gets []$
        \For{$i = 0$ to $T-1$}
            \State $a_i\gets (\Delta \mathbf{v}_i, \Delta \mathbf{t}_i,\Delta \mathcal{X}_i) \sim \pi_{\theta_{\text{old}}}(\cdot | s_i)$
            \State $s_{i+1} \gets (\mathbf{v}_i + \Delta \mathbf{v}_i, \mathbf{t}_i + \Delta \mathbf{t}_i, \mathcal{X}_i + \Delta \mathcal{X}_i)$ 
                \State $I^* = \max_{P_{\mathbf{X}}: \mathbb{E}_{\mathbf{X}}(\|\mathbf{X}\|_2^2)\leq P_T} I(\mathbf{X}; \widehat{\mathbf{W}} | s_{i+1})$
                \State $r_i \leftarrow I^* - \lambda_1 \mathcal{L}_{\text{power}}$
            \State Append $(s_i, a_i, r_i, s_{i+1})$ to $\tau_m$
            \State $s_i \leftarrow s_{i+1}$
        \EndFor
        \State Add $\tau_m$ to $\mathcal{D}$
    \EndFor
    \For{epoch = 1 to K}
            \State $\hat{g}_{\text{policy}} \gets \frac{1}{|\mathcal{D}|} \sum_{\tau \in \mathcal{D}} \sum_{i=0}^{|\tau|-1} \gamma^i \nabla_{\theta} \log \pi_{\theta}(a_i | s_i) r_i$
            \State $\hat{g}_{\text{KL}} \gets \beta \nabla_{\theta} \left( \frac{1}{|\mathcal{D}|} \sum_{\tau \in \mathcal{D}} \frac{1}{|\tau|} \sum_{i=0}^{|\tau|-1} \text{KL}(\pi_{\theta_{\text{old}}} \,\|\, \pi_\theta) \right)$
            \State $\hat{g} \gets \hat{g}_{\text{policy}} - \hat{g}_{\text{KL}}$
        \State $\theta \leftarrow \theta + \alpha \hat{g}$
    \EndFor
\EndFor
\State \textbf{return} $\theta$
\end{algorithmic}
\end{algorithm}

\subsection{Reward Computation via Mutual Information Estimation}

A critical component of the proposed RL framework is the calculation of the reward $R_i$ assigned after transitioning to state $S_{i+1} = (\mathbf{v}_{i+1}, \mathbf{t}_{i+1}, \mathcal{X}_{i+1})$ under the sampled $\text{SNR}_m$. As defined previously, this reward primarily depends on estimating the maximum achievable mutual information $I^* = \max_{P_{\mathbf{X}}} I(\mathbf{X}; \widehat{\mathbf{W}} | S_{i+1}, \text{SNR}_m)$, subject to the average power constraint $\mathbb{E}_{\mathbf{X}}[\|\mathbf{X}\|_2^2]\leq P_T$. 

The standard method for capacity estimation involves the Blahut-Arimoto (BA) algorithm \cite{blahut1972computation}, which iteratively computes the capacity and the optimal input distribution $P_{\mathbf{X}}^*$ for a given channel. While the BA algorithm can accurately determine $I^*$ for a fixed configuration $S_{i+1}$, it presents two major drawbacks for our RL setting: firstly, it is computationally intensive, needing to be run within every reward calculation step; secondly, and more critically, the BA algorithm's iterative nature make it extremely difficult, if not impossible, to directly backpropagate gradients through it. This hinders end-to-end training, as we cannot easily compute the gradient of the reward $R_i$ (derived from $I^*$) with respect to the policy parameters $\theta$ that produced $S_{i+1}$.

An alternative is to use neural network-based mutual information estimators such as CORTICAL \cite{letizia2022}.
The primary advantage of CORTICAL in our context is its inherent differentiability. This allows the gradient of the estimated MI (used in $R_i$) to flow back through the estimator and the channel model to update the policy parameters $\theta$ via standard backpropagation. Consequently, in our implementation, we have used CORTICAL to estimate channel capacity. The CORTICAL neural network and the policy network are trained iteratively in our training procedure.

\begin{figure*}[t] 
    \centering    \includegraphics[width=0.3\textwidth]{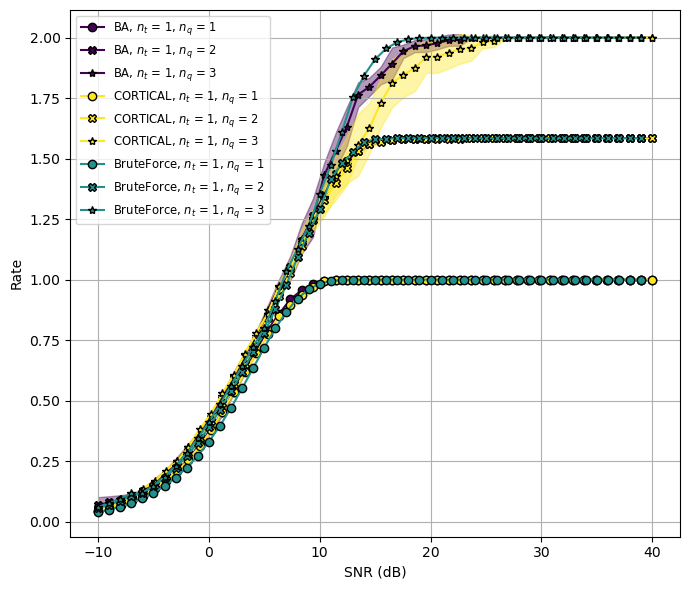}
    \includegraphics[width=0.3\textwidth]{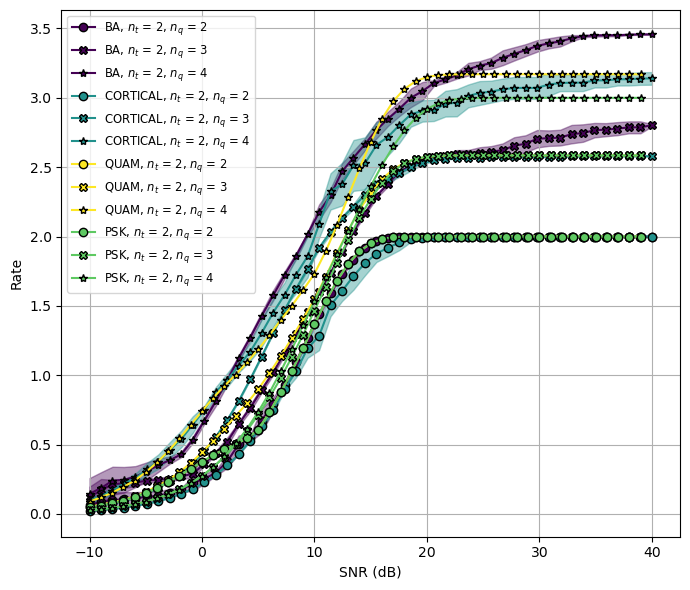}
    \includegraphics[width=0.3\textwidth]{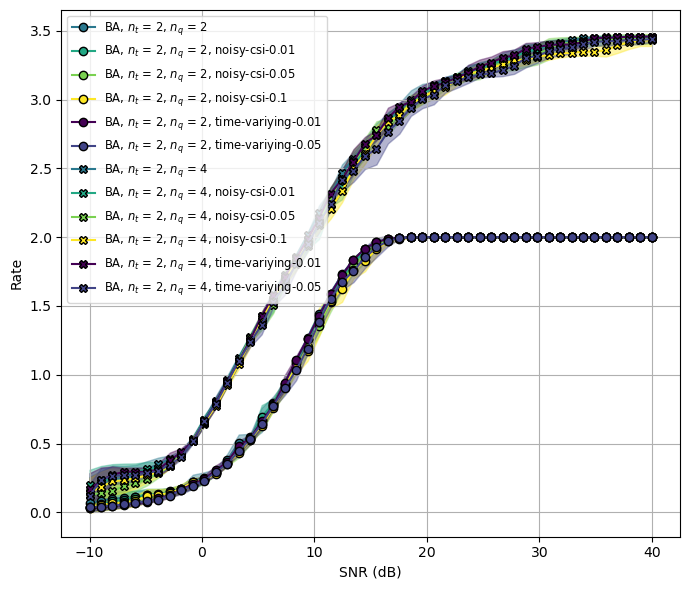}
    \vspace{-.2in}
    \caption{Capacity as a function of SNR in various communication scenarios.}
    \label{fig:snr-rate}
    \vspace{-.2in}
\end{figure*}

\section{Theoretical Analysis}
In the previous sections, we have proposed an RL mechanism to compute the parameters of the optimal quantization constellation, i.e., the hybrid beamforming matrix $\mathbf{v}$, threshold vector $\mathbf{t}$, and reconstruction points $\mathcal{X}$ maximizing channel capacity. In this section, we provide a theoretical justification for this framework, by showing that RL-based approaches converge to the optimal solution in the MIMO communication scenario under consideration, and hence their use is justified.

Let us recall the Bellman equation associated with a fixed policy $\pi$ in an MDP (e.g., \cite{sutton2018reinforcement}):
\begin{align*}
    V^{\pi}(\!s\!)\! =\!  \!\mathbb{E}_\pi(R(s,\pi(s))\!+\!\gamma\!\! \sum_{s'\in \mathcal{S}}\!\! P_{S_i|S_{i-1},A_{i-1}}\!(s'|s,\pi(s)))V^{\pi}\!(s').
\end{align*}
The optimal policy is defined as:
\begin{align*}
\pi^*(s) = \argmax_{\pi}  V^{\pi}(s),\qquad  s\in \mathcal{S}.
\end{align*}
The policy gradients methods considered in this work finds an approximate solution to the \textit{greedy} policy maximizing the state-value function $Q^\pi(s)$. We show that this greedy policy converges to the optimal policy in the MIMO communication scenario under consideration.
To this end, we introduce a \textit{truncated and discretized}  MDP, which restricts the state space to a bounded, discrete hypercube. For a sufficiently large positive constant $m > 0$ and a step size $\delta > 0$, the truncated and discretized MDP is $\mathcal{M}_{m,\delta} = (\mathcal{S}_{m,\delta}, \mathcal{A}, P_{m,\delta}, R_{m,\delta}, \gamma)$, where $\mathcal{S}_{m,\delta}$ is a finite subset of the original state space $\mathcal{S}$, where each component of the state vector is confined to the interval $[-m, m]$ and takes values in the discrete set $[-m,m]\cap \mathbb{Z}[\delta]$. 
Given a current state $s_i \in \mathcal{S}_{m,\delta}$ and an action $a_i \sim \pi_{\theta}(\cdot | s_i)$, an intermediate state is computed as
\[
\widetilde{s}_{i+1} = (\widetilde{\mathbf{v}}_{i+1}, \widetilde{\mathbf{t}}_{i+1}, \widetilde{\mathcal{X}}_{i+1}) = s_i + a_i.
\]
This intermediate state is then projected onto the discrete hypercube using the operator $\text{Proj}_{m,\delta}(\cdot)$, which maps each scalar element of $\widetilde{\mathbf{v}}_{i+1}$, $\widetilde{\mathbf{t}}_{i+1}$, and $\widetilde{\mathcal{X}}_{i+1}$ to the nearest value in $[-m,m]\cap \mathbb{Z}[\delta]$, yielding $s_{i+1} = \text{Proj}_{m,\delta}(\widetilde{s}_{i+1})$. 
The reward function $R_{m,\delta}$ is evaluated using the projected state $s_{i+1}$.
The action space $\mathcal{A}$, policy $\pi_\theta$, and discount factor $\gamma \in [0, 1]$ remain unchanged. Since $\mathcal{S}_{m,\delta}$ is finite, $\mathcal{A}$ is unchanged, and $P_{m,\delta}$ and $R_{m,\delta}$ are well-defined, it is straightforward to verify that this truncated and discretized decision process is an MDP. 

\begin{Theorem}
\label{th:1}
Given $m,\delta>0$, consider the  MDP $(\mathcal{S}_{m,\delta}, \mathcal{A}, P_{m,\delta}, R_{m,\delta}, \gamma)$, and define the greedy policy $\pi_{m,\delta}$: 
\begin{align*}
    &\pi_{m,\delta,i}(s) = \argmax_{a\in \mathcal{A}} Q^{\pi_{m,\delta,i-1}}(s,a),
  \quad 
    \pi_{m,\delta}=\lim_{i\to\infty} \pi_{m,\delta,i}.
\end{align*}
Then, 
\begin{align*}
   \lim_{\delta\to 0}V^{\pi_{m,\delta}}(s) \leq V^{\pi^*}(s)\leq \lim_{\delta\to 0}V^{\pi_{m,\delta}}(s)+O(\frac{1}{m^2})
\end{align*}
In particular, $\pi_{m,\delta}$ converges to $\pi^*$ as $m$ becomes asymptotically large and $\delta$ becomes asymptotically small. 
\end{Theorem}
The proof is provided in the Appendix.
\begin{figure}[h] 
    \centering
    \includegraphics[width=0.40\textwidth]{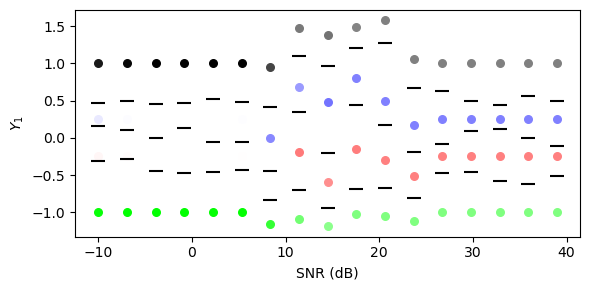}
    \vspace{-.2in}
    \caption{Input values (points) and threshold values (lines) for SISO with $n_q=3$. Point brightness indicates input probability.}
    \label{fig:1d-constellation}
    \vspace{-.2in}
\end{figure}
\begin{figure*}[h] 
    \centering
    \includegraphics[width=0.95\textwidth]{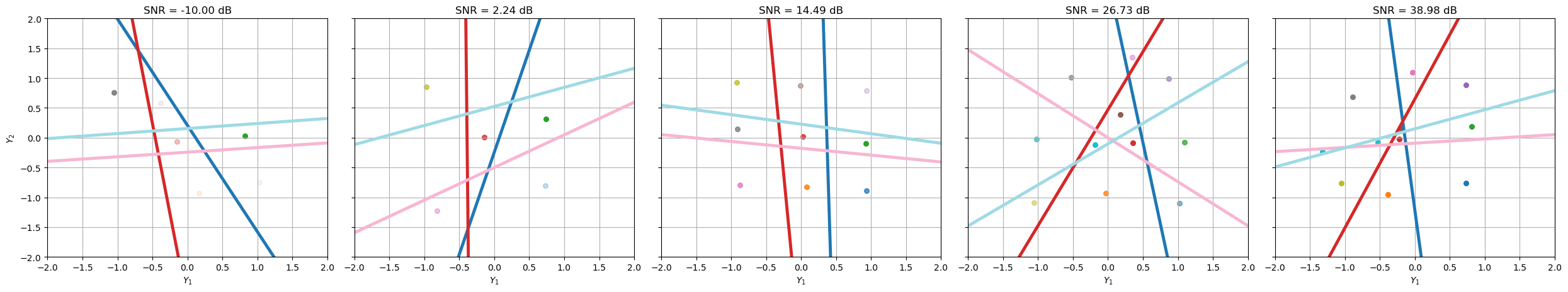}
    \vspace{-.2in}
    \caption{Learned constellations in MIMO with $n_r=2$ and $n_q=4$. Point brightness indicates input probability.}
    \label{fig:2d-constellation}
\vspace{-.2in}
\end{figure*}

\section{Numerical Simulations}
To demonstrate the near-optimal performance of our proposed RL mechanism, we conducted numerical simulations across various communication scenarios. We compared the achieved channel capacity with the results obtained through brute-force optimization, where feasible.
\\\textbf{Policies and Sub-Policies:}
We train two policy networks, one for $\mathcal{X}$, and the other for $(\mathbf{v},\mathbf{t})$.
In our implementation, the underlying neural network for each policy has three layers, with hidden-layer width equal to 192 and 384\footnote{For two-dimensional constellations, we use a slightly larger network. Details can be found on \url{https://github.com/mtalonso-research/RL_MIMO.git}.}.
Each network consists of three sub-networks (sub-policies) trained separately for i) low SNR: [-10,0], ii) mid SNR: [0,10], and iii) high SNR: [10,40].  
\\\textbf{Simulation Setup:}
We initialized 30 distinct environments, with $(\mathbf{v}, \mathbf{t})$ symmetrically placed around the origin and $\mathcal{X}$ at the centroids of the resulting regions. Training spanned 5 episodes with a learning rate of 0.001, using a unified policy across 10 environments, a maximum of 2000 steps per episode, and early stopping after 100 steps without improvement.
\\\textbf{Testing and Inference:} For each environment, the policy runs for a maximum of 1000 steps, with early stopping enabled after 100 consecutive steps without improvement. We perform ten inference runs on each SNR value in the interval $[-10,40]$, and report the mean and standard deviation across these runs in each scenario.
\\\textbf{Experiment 1: Single-Input Single-Output (SISO)} We simulated SISO scenarios with SNRs in [-10, 40] dB and one-bit ADCs ($n_q \in {1, 2, 3}$). Figure \ref{fig:snr-rate}(a) compares channel capacities for: (i) our RL method using the Blahut-Arimoto (BA) algorithm for $P_X$, (ii) our RL method with a trained CORTICAL network for $P_X$, and (iii) brute-force optimization (optimal $(\mathbf{v}, \mathbf{t}, \mathcal{X})$). Shaded regions indicate standard deviations across ten runs. The RL policies closely match the brute-force baseline, with BA-trained policies showing lower variance. Figure \ref{fig:1d-constellation} illustrates the learned quantization constellations, revealing two mass points at low SNR, increasing to four at higher SNR, consistent with prior findings on optimal constellations in quantized MIMO systems \cite{dytso2018discrete}.
\\\textbf{Experiment 2: Multiple-Input Multiple-Output (MIMO)}
For MIMO with two receive antennas ($n_r=2$), Figure \ref{fig:snr-rate}(b) compares channel capacities of RL with BA and CORTICAL against QAM and PSK baselines. The RL method significantly outperforms these baselines. Brute-force optimization is infeasible due to high-dimensional complexity. Figure \ref{fig:2d-constellation} shows learned constellations, with more mass points activated as SNR increases. At SNR = 14.49 dB, the constellation resembles QAM, transitioning to a general position constellation at higher SNR, as predicted in \cite{khalili2021mimo}.
\\\textbf{Experiment 3: Noisy Channel State Information (CSI)}
We tested BA-trained policies under noisy CSI, modeled as zero-mean Gaussian noise with variances of 0.01, 0.05, and 0.1, and non-stationary time-varying channels. Figure \ref{fig:snr-rate}(c) shows that channel capacity remains robust across these noise levels.
\section{Conclusion}
We have introduced an RL framework for receiver-side optimization in quantized MIMO settings. The proposed approach utilizes neural network-based
mutual information estimators for reward calculation. It was shown through extensive simulations that the approach achieves near optimal performance, in terms of resulting channel capacity, and 
is
robust to dynamic channel statistics and noisy CSI estimates.

\bibliographystyle{IEEEtran}
\bibliography{References}
\begin{appendix}
\textit{Proof of Theorem \ref{th:1}:}
\label{App:th:1}
We provide an outline of the proof. 
    It is well-known that the greedy policy approaches the optimal policy in finite MDPs (e.g., \cite{sutton2018reinforcement}). Thus, it suffices to show that the maximum reward (channel capacity) achieved using the  greedy policy for the truncated and discretized MDP approaches that of the original MDP as $m\to \infty$ and $\delta\to 0$. That is, we need to show that:
    \begin{align*}
    \lim_{\delta\to 0}C_{m,\delta}\leq C\leq \lim_{\delta\to 0} C_{m,\delta}+O(\frac{1}{m^2}), 
    \end{align*}
    where
    \begin{align*}
        & C=\max_{(\mathbf{v},\mathbf{t}, \mathcal{X})}\max_{P_{\mathbf{X}}} I(\mathbf{X};\mathbf{\widehat{W}}),\\
        &C_{m,\delta}=\max_{(\mathbf{v},\mathbf{t}, \mathcal{X})\in ([-m,m]\cap \mathbb{Z}[\delta])^{n_qn_r+n_q(\ell-1)+n_r\zeta}}\max_{P_{\mathbf{X}}} I(\mathbf{X};\mathbf{\widehat{W}}).
    \end{align*}
    The lower-bound follows from \cite[Lemmas 2 and 3]{shirani2023structured}. To prove the upper-bound, as an intermediate step, let us define:
    \begin{align*}
C_m=\max_{(\mathbf{v},\mathbf{t}, \mathcal{X})\in [-m,m]^{n_qn_r+n_q(\ell-1)+n_r\zeta}}\max_{P_{\mathbf{X}}} I(\mathbf{X};\mathbf{\widehat{W}}).
    \end{align*}
    Then, from \cite[Lemma 2]{shirani2023structured}, we have:
    \begin{align*}
        C_m \leq\lim_{\delta\to 0}C_{m,\delta}.
    \end{align*}
    So, it suffices to show that:
    \begin{align*}
        C_m=C+O(\frac{1}{m^2}). 
    \end{align*}
    Let $(\mathbf{v}^*,\mathbf{t}^*,\mathcal{X}^*, P^*_{\mathbf{X}})$ represent the parameters which achieve $C$. Let $\widehat{C}$ be value of $I(\mathbf{X};\widehat{\mathbf{W}})$ evaluated over $(\mathbf{v}^*,\mathbf{t}^*,\mathcal{X}^*, P^*_{\mathbf{X}|\mathbf{X}\in [-m,m]^{n_r}})$. Note that due to the power constraint $\mathbb{E}(\|\mathbf{X}\|^2_2)\leq P_T$, we have $P(\mathbf{X}\notin  [-m,m]^{n_r})= O(\frac{1}{m^2})$. So,
    \begin{align*}
        &C_m
        = \max_{(\mathbf{v},\mathbf{t}, \mathcal{X})\in [-m,m]^{n_qn_r+n_q(\ell-1)+n_r\zeta}}\max_{P_{\mathbf{X}}}I(\mathbf{X};\mathbf{\widehat{W}})
        \\&\geq  I(\mathbf{X};\widehat{\mathbf{W}}|\mathbf{v}^*,\mathbf{t}^*,\mathcal{X}^*, P^*_{\mathbf{X}|\mathbf{X}\in [-m,m]^{n_r}})
        \\&\geq P(\mathbf{X}\in  [-m,m]^{n_r})C -P(\mathbf{X}\notin  [-m,m]^{n_r})\log(\ell^{n_q})
        \\& = C+O(\frac{1}{m^2}).
    \end{align*}
    This completes the proof. 
    
\end{appendix}
\end{document}